# Asymptotic Granularity Reduction and Its Application[*]

Shenghui Su[1], Shuwang Lü[2], and Xiubin Fan[3]

[1] College of Computer, Beijing University of Technology, Beijing 100124, P.R.China
[2] Graduate School, Chinese Academy of Sciences, Beijing 100039, P.R.China
[3] Institute of Software, Chinese Academy of Sciences, Beijing 100080, P.R.China

**Abstract**: It is well known that the inverse function of $y = x$ with the derivative $y' = 1$ is $x = y$, the inverse function of $y = c$ with the derivative $y' = 0$ is nonexistent, and so on. Hence, on the assumption that the noninvertibility of the univariate increasing function $y = f(x)$ with $x > 0$ is in direct proportion to the growth rate reflected by its derivative, the authors put forward a method of comparing difficulties in inverting two functions on a continuous or discrete interval called asymptotic granularity reduction (AGR) which integrates asymptotic analysis with logarithmic granularities, and is an extension and a complement to polynomial time (Turing) reduction (PTR). Prove by AGR that inverting $y \equiv x^x \pmod{p}$ is computationally harder than inverting $y \equiv g^x \pmod{p}$, and inverting $y \equiv g^{x^n} \pmod{p}$ is computationally equivalent to inverting $y \equiv g^x \pmod{p}$, which are compatible with the results from PTR. Besides, apply AGR to the comparison of inverting $y \equiv x^n \pmod{p}$ with $y \equiv g^x \pmod{p}$, $y \equiv g^{g_1^x} \pmod{p}$ with $y \equiv g^x \pmod{p}$, and $y \equiv x^n + x + 1 \pmod{p}$ with $y \equiv x^n \pmod{p}$ in difficulty, and observe that the results are consistent with existing facts, which further illustrates that AGR is suitable for comparison of inversion problems in difficulty. Last, prove by AGR that inverting $y \equiv x^n g^x \pmod{p}$ is computationally equivalent to inverting $y \equiv g^x \pmod{p}$ when PTR cannot be utilized expediently. AGR with the assumption partitions the complexities of problems more detailedly, and finds out some new evidence for the securities of cryptosystems.

## 1 Introduction

Cryptography is the foundation stone of trusted computing and information security. In public key cryptosystems, the security of a data encryption or digital signature scheme is based on an intractable computational problem which cannot be solved in polynomial or subexponential time. For instance, the RSA scheme is based on the integer factorization problem (IFP) [1], and the ElGamal scheme is based on the discrete logarithm problem (DLP) [2].

If a scheme or protocol is proven secure on the assumption that IFP and DLP cannot be solved in polynomial time, it is said to be secure in the standard model. Generally, security proofs are difficult to achieve in the standard model, and thus, sometimes cryptographic primitives are idealized ─ a hash function is regarded as truly stochastic in the random oracle model for example [3].

Polynomial time Turing reduction, in brief polynomial time reduction (PTR), is usually employed to compare the complexities or difficulties of two computational problems [4][5]. Obviously, results from PTR provide some evidence for the securities of cryptosystems, but not any two computational problems can be compared suitably through PTR.

The complexity or difficulty of a computational problem is related to the time complexity of the fastest algorithm (if existent) for solving the problem. Complexities of problems may be coarsely partitioned into three levels: computable in polynomial time, computable in superpolynomial time ─ in subexponential or exponential time for example ─ and undecidable, namely unsolvable through an algorithm [4]. A problem belongs to the class P if it can be solved on a deterministic Turing machine in polynomial time, or the class NP if it can be solved on a nondeterministic Turing machine in polynomial time [6]. A P problem is regarded as tractable since a polynomial time algorithm for solving it can be found, and an NP problem is regarded as intractable or hard since a polynomial time algorithm for solving it is not found yet [7]. It is an open and hot topic at present whether P ≠ NP or not.

There are certain problems in NP whose individual complexity is related to that of the entire class. If a polynomial time algorithm exists for any of these problems, all problems in NP would be solvable in polynomial time. These problems are called NP-complete [7]. That is to say, suppose that $A$ is NP-complete, and then $A \in$ P if and only if P = NP [6].

---

[*] Manuscript received on 01 Jun 2011, and last revised 11 Dec 2011. It occurs in TCS (vol. 412(39), Sep 2011, pp. 5374-5386).
This work is supported by MOST with Project 2007CB311100 and 2009AA01Z441. Corresponding email: reesse@126.com.

Based on an assumption, a new method of comparing complexities of two computational problems which combines asymptotic analysis with logarithmic granularities is proposed in this paper. The method possesses a theoretical value that a new measure of a relative complexity of a problem which avoids seeking of the fastest algorithm for solving the problem is given, and a practical value that some new evidence of cryptosystem security can be found through it.

Throughout the paper, unless otherwise specified, $p \neq 2$ is a large prime number, $g_0 \in \mathbb{Z}_p^*$, $g_1 \in \mathbb{Z}_{\phi(p)}^*$, $g_2 \in \mathbb{Z}_{\phi(\phi(p))}^*$ etc are some group elements ($g_i$ is a generator if it belongs to a cyclic group), $n \geq 5$ and $g > 1$ are either two integers with $g = g_0$ in a discrete interval or two rationals in a continuous interval, $\lg x$ is the logarithm of $x$ to the base 2, $\ln x$ is the natural logarithm of $x$, $\log x$ is the logarithm of $x$ to the base $g$, the sign % denotes modular arithmetic, $\phi$ denotes an Euler phi function, $\cong$ denotes the equivalence of two limits, the signs $\mathbb{Z}$ and $\mathbb{R}$ represent the sets of integral and real numbers respectively, and the time complexity of a algorithm is measured in bit operations.

## 2  Polynomial Time Reduction and Asymptotic Granularity Reduction

Let $c > 1$ be any constant, and $x$ be an input of an algorithm. Then the time complexity of the fastest algorithm (if existent) for solving a problem may be logarithmic in $x$ — $O(\lg x)$, linear — $O(x)$, polynomial — $O(x^c)$, subexponential — $O(c^{o(1)x})$ with $0 < o(1) < 1$, exponential — $O(c^x)$, or factorial — $O(x!)$ for example.

If the time complexities of the two fastest algorithms respectively for solving the problems $A$ and $B$ are on the same level, the difficulty of $A$ is said to be equivalent to that of $B$. If the time complexity of the fastest algorithm for $A$ is lower than that of the fastest algorithm for solving $B$, the difficulty of $A$ is said to be less than that of $B$. For example, if the time complexities of the two fastest algorithms respectively for $A$ and $B$ are linear and polynomial, the difficulty of $A$ is said to be less than that of $B$ although both $A$ and $B$ are efficiently computable. Thus, there exists a partial order relation among the difficulties of problems [8].

In this section, we will give some definitions, conceps, and explanations relevant to polynomial time reduction and asymptotic granularity reduction.

### 2.1  Polynomial Time Reduction and Asymptotic Security

To compare the complexities or difficulties of two computational problems described with a univariate function, PTR is usually utilized [5].

**Definition 1**: Let $A$ and $B$ be two computational problems. $A$ is said to reduce to $B$ in polynomial time, written as $A \leq_P B$, if there is an algorithm for solving $A$ which calls, as a subroutine, a hypothetical algorithm for solving $B$, and runs in polynomial time excluding the running time of the algorithm for solving $B$.

The hypothetical algorithm for solving $B$ is called an oracle. It is not difficult to understand that no matter what the running time of the oracle is, it does not influence the result of the comparison.

$A \leq_P B$ means that the difficulty of $A$ is not greater than that of $B$, namely the complexity of an algorithm for solving $A$ is not greater than that of an algorithm for solving $B$ when all polynomial times are treated as the same. Concretely speaking, if $A$ is unsolvable in polynomial or subexponential time, $B$ is also unsolvable in polynomial or subexponential time; and if $B$ is solvable in polynomial or subexponential time, $A$ is also solvable in polynomial or subexponential time.

**Definition 2**: Let $A$ and $B$ be two computational problems. If $A \leq_P B$ and $B \leq_P A$, then $A$ and $B$ are said to be computationally equivalent, written as $A =_P B$.

Definition 1 and 2 suggest polynomial time reduction, a reductive proof method. Provable security by PTR is substantially relative and asymptotic just as a one-way function is. Relative security implies that the security of a cryptosystem based on a problem is comparative, but not absolute. Asymptotic security implies that even if a cryptosystem based on a problem is proven to be secure, it is practically secure only on condition that the dominant parameter is large enough. Of course, to different problems, the asymptotic tendencies are distinct.

Naturally, we will consider $A <_P B$. Perhaps it is easy to give the definition of $A <_P B$ theoretically, but it is not easy to give the formal proof of $A <_P B$ practically. See the following example.

**Definition 3**: Let $n \geq 80$ be an integer, $p$ a prime with $\lg p \leq n$, and $\{C_1, \ldots, C_n\}$ a sequence with every $C_i < p$, then seeking the nonzero binary string $b_1 \ldots b_n$ from known $\bar{G} \equiv \prod_{i=1}^{n} C_i^{b_i}$ (% $p$) (particularly, $\bar{G} \equiv \prod_{i=1}^{n} C_i^{b_i 2^{i-1}}$ (% $p$) when $C_1 = \ldots = C_n$) is called the subset product problem, in brief SPP.

Let $\bar{O}_p(\bar{G}, \{C_1, \ldots, C_n\})$ be an oracle on solving $\bar{G} \equiv \prod_{i=1}^{n} C_i^{b_i}$ (% $p$) for $b_1 \ldots b_n$, and DLP be of solving $y \equiv g^x$ (% $M$) for $x$.

Then, by calling $\bar{O}_p(y, \{g, \ldots, g\})$, $x$ can be found, and according to definition 1, there is

$$\text{DLP} \leq_P \text{SPP}.$$

Moreover, if $C_1 = \ldots = C_n = C$, inverting $\bar{G} \equiv \prod_{i=1}^{n} C_i^{b_i} \equiv C^z$ (% $p$) for $z$ is equivalent to DLP, where $z = \sum_{i=1}^{n} b_i 2^{i-1}$. Nevertheless in practice, we may make $C_1, \ldots, C_n$ pairwise distinct by checking $C_1, \ldots, C_n$ in advance so that the condition $C_1 = \ldots = C_n$ will not occur forever, namely the case of SPP being equivalent to DLP will not occur forever.

Additionally, presuppose that DLP can be solved through an oracle.

Let $g$ be a generator of $(\mathbb{Z}_p^*, \cdot)$, $C_1 \equiv g^{u_1}$ (% $p$), …, $C_n \equiv g^{u_n}$ (% $p$), $\bar{G} \equiv g^v$ (% $p$).

Then, solving $\bar{G} \equiv \prod_{i=1}^{n} C_i^{b_i}$ (% $p$) is equivalent to solving

$$b_1 u_1 + \ldots + b_n u_n \equiv v \text{ (% } p - 1 \text{)},$$

which is a subset sum problem (SSP). Due to $\lg p < n$, the related knapsack density $D = n / \lg p$ is greater than 1.

It is well known that SSP is NP-complete (in its feasibility recognition form) [6], and difficult to compute. Especially, when the knapsack density $D$ is greater than 1, SSP is resistant to $L^3$ lattice base reduction attack [9].

The above two pieces of evidence incline us to believe that SPP is harder than DLP in the same prime field, namely SPP cannot be solved in DLP subexponential time, and yet it is not easy to give a satisfying and convincing proof of DLP $<_P$ SPP because the proof will indicate that P $\neq$ NP holds.

## 2.2 Big-*O* Notation and Asymptotic Analysis

In theory of computation, the notation big-*O* frequently occurs.

Because the exact time complexity, namely running time of an algorithm is a complicated expression in general, we usually just estimate it. In a convenient form of estimation called asymptotic analysis, we seek to know the time complexity of the algorithm when it is run on large input values [10].

We do so by considering only the highest degree term of the expression for the time complexity but neglecting both the coefficient of that term and any lower degree terms because the highest degree term dominates the other terms on large input values.

For example, assume that the time complexity of an algorithm is $f(x) = 5x^4 + 11x^3 + 31x^2 + 7x + 23$ which has five terms. Neglecting the coefficient 5 of the highest degree term $5x^4$ and the other lower degree terms, we say that $f$ is asymptotically at most $x^4$, namely there is an asymptotic expression $f(x) = O(x^4)$ since

$$\lim_{x \to \infty} (5x^4 + 11x^3 + 31x^2 + 7x + 23) / x^4 = 5$$

is a constant, where $x$ need not be continuous.

## 2.3 Some Facts and Phenomena

It is well understood that the growth rate of a univariate function is decided by its derivative. It can be observed that the inverse of the function $y = x$ with the derivative $y' = 1$ is $x = y$, and the inverse of the function $y = c$ with the derivative $y' = 0$ is nonexistent, which hints that the one-wayness or noninvertibility of an increasing real function $y = f(x)$ with $x \in (0, \infty)$ should be in direct proportion to its growth rate, namely difficulty in inverting $y = f(x)$ should be reflected by its derivative.

### 2.3.1 Difficulties of Several Inversion Problems on a Continuous Interval

***Definition 4***: Let $y = f(x)$ is an increasing real function with $x \in (0, \infty)$ and $y \in \mathbb{R}$, then inverting $y = f(x)$, namely computing $x$ from known $y$ is called an inversion problem.

Of course, an inversion problem is a computational problem.

Observe the following facts.

Evidently, for the identity function $y = x$, its inverse can be found in linear time.

For the reciprocal function $y = -1 / x$, its inverse is involved in an infinite circulating decimal of which the repeated digits can be found in quadratic polynomial time.

For the power function $y = x^n$ or $y = c_n x^n + c_{n-1} x^{n-1} + \ldots + c_0$ with $n \geq 5$, its inverse is involved in an irrational root, namely an algebraic irrational [11], and cannot be found in polynomial time when its precision approaches infinity. However, we may judge that the running time of an oracle on inverting $y = x^n$ or $y = c_n x^n + c_{n-1} x^{n-1} + \ldots + c_0$ is relevant to seeking an algebraic irrational.

For the exponential function $y = g^x$, its inverse is involved in a transcendental irrational (not an algebraic irrational) [11], and cannot be found in polynomial time when its precision approaches infinity. However, we may judge that the running time of an oracle on inverting $y = g^x$ is relevant to seeking a transcendental irrational.

For the power exponential function $y = x^x$, its inverse cannot be found in polynomial time when its precision approaches infinity, where $x = \log y / \log x$ is called a transcendental logarithm. However, we may judge that the running time of an oracle on inverting $y = x^x$ is relevant to seeking a transcendental logarithm.

It should be noted that when the precision approaches infinity, the attempt to let $y = g^z$, $x = g^w$, and further $z = w g^w$ are incorrect because obtaining $z$ of infinite precision is generally infeasible no matter what value $g$ is, which make it clear that inverting $y = x^x$ cannot be reduced to inverting $y = g^x$.

Therefore, the above facts illustrate that in general, inverting $y = x^x$ is harder than inverting $y = g^x$, inverting $y = g^x$ is harder than inverting $y = x^n$, inverting $y = x^n$ is harder than inverting $y = -1/x$, and inverting $y = -1/x$ is harder than inverting $y = x$ in computational complexity.

Let $\bar{I}(y = f(x))$ represent the difficulty in solving $y = f(x)$ for $x$, namely inverting $y = f(x)$ [4]. Then, according to the above facts, there is

$$\bar{I}(y = x) < \bar{I}(y = -1/x) < \bar{I}(y = x^n) < \bar{I}(y = g^x) < \bar{I}(y = x^x).$$

Further observation.

Let

$$f_1(x) = y = x, f_2(x) = y = -1/x, f_3(x) = y = x^n, f_4(x) = y = g^x, \text{ and } f_5(x) = y = x^x,$$

then their derivatives are separately

$$f_1'(x) = 1, f_2'(x) = 1/x^2, f_3'(x) = n x^{n-1}, f_4'(x) = g^x \ln g, \text{ and } f_5'(x) = x^x (\ln x + 1).$$

An interesting phenomenon may be discovered.

For $f_1'(x)$ and $f_2'(x)$, there is

$$\lim_{x \to \infty} \log f_1'(x) / \log f_2'(x) = \lim_{x \to \infty} \log (1) / \log (1/x^2)$$
$$= \lim_{x \to \infty} 0 / (-2 \log x)$$
$$= 0,$$

which is consistent with the fact that inverting $y = -1/x$ is generally infeasible in linear time, namely inverting $y = -1/x$ is generally harder than inverting $y = x$.

For $f_2'(x)$ and $f_3'(x)$, there is

$$\lim_{x \to \infty} \log f_2'(x) / \log f_3'(x) = \lim_{x \to \infty} \log (1/x^2) / \log (n x^{n-1})$$
$$= \lim_{x \to \infty} (-2 \log x) / (\log n + (n-1) \log x)$$
$$= 0$$

as $n \geq 5$, which is consistent with the fact that inverting $y = x^n$ is generally infeasible in polynomial time, namely inverting $y = x^n$ is generally harder than inverting $y = -1/x$.

For $f_3'(x)$ and $f_4'(x)$, there is

$$\lim_{x \to \infty} \log f_3'(x) / \log f_4'(x) = \lim_{x \to \infty} \log (n x^{n-1}) / \log (g^x \ln g)$$
$$= \lim_{x \to \infty} (\log n + (n-1) \log x) / (x + \log \ln g)$$
$$= 0,$$

which is consistent with the fact that inverting $y = g^x$ is generally infeasible in time of seeking an algebraic irrational, namely inverting $y = g^x$ is generally harder than inverting $y = x^n$.

For $f_4'(x)$ and $f_5'(x)$, there is

$$\lim_{x \to \infty} \log f_4'(x) / \log f_5'(x) = \lim_{x \to \infty} \log (g^x \ln g) / \log (x^x (\ln x + 1))$$
$$= \lim_{x \to \infty} (x + \log \ln g) / (x \log x + \log (\ln x + 1))$$
$$= 0,$$

which is consistent with the fact that inverting $y = x^x$ is generally infeasible in time of seeking a transcendental irrational, namely inverting $y = x^x$ is generally harder than inverting $y = g^x$.

### 2.3.2 Difficulties in Inverting Inverse Functions

It is mentioned in the above paragraphs that the growth rate of a function $y = f(x)$ is reflected by its derivative

$$y' = \lim_{\Delta x \to \infty} \Delta y / \Delta x.$$

Then, according to the calculus theory, the derivative of a corresponding inverse function is

$$x' = 1 / y' = \lim_{\Delta y \to \infty} \Delta x / \Delta y.$$

We know that the inverse of the inverse of a function is itself.

In fact, for $y = x^n$, $y = g^x$, and $y = x^x$ with $x \in (0, \infty)$, difficulties in computing $y$ from $x$ are equivalent, and computing $y$ from $x$ is relatively easier than computing $x$ from $y$.

Now, let the inverse functions of $y = x^n$, $y = g^x$, and $y = x^x$ be respectively

$$x = \lambda_1(y), x = \lambda_2(y), \text{ and } x = \lambda_3(y).$$

Then, there are

$$\lambda_1'(y) = y^{1/n} / (n\,y),\ \lambda_2'(y) = 1 / (y \ln g),\ \text{and } \lambda_3'(y) = 1 / (y(\ln x + 1)).$$

Clearly every derivative is a decimal less than 1 when $y > 1$, which illuminates that the growth of every inverse function is slow.

Notice that when we compute $y$ from the inverse function $x = \lambda_1(y)$, $x = \lambda_2(y)$, or $x = \lambda_3(y)$, known $x$ is a real with a finite precision in practice.

Another interesting phenomenon may be discovered.

For $\lambda_1'(y)$ and $\lambda_2'(y)$, there is

$$\lim_{y \to \infty} \log \lambda_1'(y) / \log \lambda_2'(y) = \lim_{y \to \infty} \log(y^{1/n} / (n\,y)) / \log(1 / (y \ln g))$$
$$= \lim_{y \to \infty} (\log y + \log n - (1/n)\log y) / (\log y + \log \ln g)$$
$$= 1,$$

which indicates that computing $y$ from $y = x^n$ is equivalent to computing $y$ from $y = g^x$ in complexity, and is consistent with the fact.

For $\lambda_2'(y)$ and $\lambda_3'(y)$, due to $\lim_{y \to \infty} \log(\ln x + 1) / \log y = 0$, where $x$ satisfies $y = x^x$, there is

$$\lim_{y \to \infty} \log \lambda_3'(y) / \log \lambda_2'(y) = \lim_{y \to \infty} \log(1 / (y(\ln x + 1))) / \log(1 / (y \ln g))$$
$$= \lim_{y \to \infty} (\log y + \log(\ln x + 1)) / (\log y + \log \ln g)$$
$$= 1,$$

which indicates that computing $y$ from $y = x^x$ is equivalent to computing $y$ from $y = g^x$ in complexity, and is consistent with the fact.

### 2.4 Asymptotic Granularity Reduction ─ an Extension of Asymptotic Analysis

According to the facts and phenomena in section 2.3, we acquire the following assumption and definitions.

***Assumption 1***: Let $y = f(x)$ with $x \in (0, \infty)$ be an increasing real function. Then the noninvertibility of $y = f(x)$, namely difficulty in inverting $y = f(x)$ is directly proportional to the growth rate reflected by its derivative.

The facts and phenomena observed incline us to believe that assumption 1 is valid although the proof of it is as difficult as the proof of the Church-Turing proposition [6].

### 2.4.1 Asymptotic Granularity Reduction over a Continuous Interval

On the basis of assumption 1 and the connotation of big-$O$, we conceive and give the expression

$$\lim_{x \to \infty} \log^k f'(x) / \log^k h'(x)$$

which may be used for the comparison of difficulties in inverting the two univariate increasing functions $f(x)$ and $h(x)$, where $k \geq 0$ is an integer, and represents a logarithmic granularity, namely the number of times of logarithmic operation.

Let $\log^0 f'(x) = f'(x)$, $\log^1 f'(x) = \log_g f'(x)$, $\log^2 f'(x) = \log_g(\log_g f'(x))$ etc.

The selection of a logarithmic granularity should make the comparison of two derivatives performable and the result reasonable. In most cases, we select $k = 1$, which indicates that the bit-lengths of the two derivatives are compared when the base $g$ is equal to 2.

**Definition 5**: Let $f(x)$, $h(x)$ be two increasing functions on the continuous interval $(0, \infty)$, and their $k$-th derivatives exist respectively, where $k \geq 0$. Then the comparison of difficulties in inverting the two functions by

$$\lim_{x \to \infty} \log^k f'(x) / \log^k h'(x)$$

which combines asymptotic analysis with a logarithmic granularity is called asymptotic granularity reduction over the continuous interval, in brief AGR.

An asymptotic granularity expression

$$\lim_{x \to \infty} \log^k f'(x) / \log^k h'(x) = 0, c, \text{ or } \infty$$

means that inverting $f(x)$ is easier than, equivalent to, or harder than inverting $h(x)$ in computational complexity, where $c$ is a constant number not less than 1, or a constant interval with the lower limit 1 in some cases.

Obviously, in terms of definition 5, we can explain the realities in section 2.3 reasonably.

### 2.4.2  Asymptotic Granularity Reduction over a Discrete Interval

In cryptology, a computational problem regarded as an intractability and described with a univariate function $y \equiv f(x)$ (% $p$) is commonly one-way, which indicates that it is very intractable to invert $y \equiv f(x)$ (% $p$), where $p$ is a positive prime. Taking $y \equiv xg^x$ (% $p$) as an example, computing $y$ from $x$ is tractable, but computing $x$ from $y$ is intractable.

We know that a discrete interval $(0, p)$ which only contains the positive integers less than $p$ is a subset of a continuous interval $(0, \infty)$. Observe the curve of $y \equiv f(x)$ (% $p$).

Let $y \equiv g^x$ (% $p$), where $g$ is a generator, and $x \in (0, \infty)$ is a real. Then its curve on the continuous interval $(0, p)$ has a sawtoothed form. There are many intersections where the line $y = y_0$ with $y_0$ being a positive integer and the curve $y \equiv g^x$ (% $p$) intersect, and among these intersections exists the unique one whose $x$-value is also an integer. Obviously, the faster the growth rate of $y \equiv f(x)$ (% $p$) is, the more the number of the intersections of $y \equiv f(x)$ (% $p$) and $y = y_0$ is, and the harder finding the point with an integral $x$-value is.

Hence, when we consider $y \equiv f(x)$ (% $p$) from mathematical analysis, the interval $(0, p)$ may be regarded as continuous, namely the modular operation of positive reals is permitted, and only when we consider $y \equiv f(x)$ (% $p$) from number theory or finite field theory, is the interval $(0, p)$ regarded as discrete. Such an idea is not fresh in cryptoanalysis. For instance, the accumulation points of minima in a continuous interval are used to attack an MH private key [12].

In this way, if $y = f(x)$ is increasing, $y \equiv f(x)$ (% $p$) is also regarded as increasing although not strictly increasing, and moreover if the $k$-th derivative of $y = f(x)$ exists, the $k$-th derivative of $y \equiv f(x)$ (% $p$) is also regarded as existing on all the points of the curve except the limited discontinuities, which means that AGR may be extended to a discrete interval.

It should be noticed that on a discrete interval, we define $\log f'(x) = \log^1 f'(x) = \log_{g_0} f'(x)$, $\log^2 f'(x) = \log_{g_1}(\log_{g_0} f'(x))$, $\log^3 f'(x) = \log_{g_2}(\log_{g_1}(\log_{g_0} f'(x)))$ etc, and in addition, because it is impossible that the tendency $x \to \infty$ occurs in practice, we substitute '$x \to \infty$' with '$x \to \phi^k(p)$ & $\phi^k(p) \to \infty$', where $k \geq 0$ is an integer, & denotes 'and', and $\phi^0(p) = p$, $\phi^1(p) = \phi(p)$, $\phi^2(p) = \phi(\phi(p))$ etc are stipulated.

Sometimes, the tendency variable may be $x^{-1}$ or $\log x$ as long as its degree is dominant, that is to say, $x^{-1} \to \phi^k(p)$ & $\phi^k(p) \to \infty$, or $\log x \to \phi^k(p)$ & $\phi^k(p) \to \infty$ possibly occurs.

**Definition 6**: Let $f(x)$ % $p$, $h(x)$ % $p$ be two increasing functions on the discrete interval $(0, p)$, and their $k$-th derivatives exist respectively, where $k \geq 0$. Then the comparison of difficulties in inverting the two modular functions by

$$\lim_{x \to \phi^k(p) \ \& \ \phi^k(p) \to \infty} \log^k f'(x) / \log^k h'(x)$$

which combines asymptotic analysis with a logarithmic granularity is called asymptotic granularity reduction over the discrete interval, in brief AGR.

An asymptotic granularity expression

$$\lim_{x \to \phi^k(p) \,\&\, \phi^k(p) \to \infty} \log^k f'(x) / \log^k h'(x) = 0, c, \text{ or } \infty$$

indicates that inverting $f(x) \% p$ is easier than, equivalent to, or harder than inverting $h(x) \% p$ in computational complexity, where $c$ is a constant number not less than 1, or a constant interval with the lower limit 1 in some cases.

Notice that in modular arithmetic, it is possible that there exists

$$\lim_{x \to \phi^k(p) \,\&\, \phi^k(p) \to \infty} (\log^k x / x) = 1, \text{ or } \lim_{x \to \phi^k(p) \,\&\, \phi^k(p) \to \infty} (x / \log^k x) = 1,$$

where $k \geq 1$.

Since DLP is secure in the standard model, and many problems need to be compared with DLP in practice, the function $y \equiv g^x (\% p)$ is often selected as $h(x)$, namely $h(x) \equiv y \equiv g^x (\% p)$. Apparently, the first logarithm of its derivative to the base $g$ is $\log h'(x) = x + \log(\ln g)$, and the constant part $\log(\ln g)$ may be neglected in asymptotic analysis.

### 2.4.3 Case of the Modulus Being a Composite Number

Let $y \equiv f(x) (\% m)$ be an increasing function, where $m = pq$ is a composite integer, including two prime factors.

Assume that $y$ is known, we need to seek $x$ satisfying $y \equiv f(x) (\% m)$.

It is well known that seeking $p$ or $q$ satisfying $m = pq$ is called the integer factorization problem (IFP). If $p$ and $q$ can be found, solving $y \equiv f(x) (\% m)$ for $x$ may be converted into solving $y \equiv f(x) (\% p)$ and $y \equiv f(x) (\% q)$ for $x$ according to $y = f(x) - kpq$.

There are the following four points which should be noticed.

Firstly, at present, IFP is regarded as being computationally equivalent to DLP owing to the number field sieve method [5].

Secondly, if $x_1$ satisfying $y \equiv f(x) (\% p)$ and $x_2$ satisfying $y \equiv f(x) (\% q)$ can be found, the original solution $x$ to $y \equiv f(x) (\% m)$ can be found from the congruence system

$$\begin{cases} x \equiv x_1 (\% p), \\ x \equiv x_2 (\% q) \end{cases}$$

in terms of the Chinese remainder theorem [13], which means that when IFP can be solved, inverting $y \equiv f(x) (\% m)$ is computationally equivalent to inverting $y \equiv f(x) (\% p)$. For example, inverting $y \equiv g^x (\% m)$ is computationally equivalent to inverting $y \equiv g^x (\% p)$.

Thirdly, through the AGR method, if inverting $y \equiv f(x) (\% p)$, where $p$ derives from $m$, is easier than or equivalent to inverting $y \equiv g^x (\% p)$, inverting $y \equiv f(x) (\% m)$ may be regarded as equivalent to $y \equiv g^x (\% p)$. If inverting $y \equiv f(x) (\% p)$ is harder than inverting $y \equiv g^x (\% p)$, inverting $y \equiv f(x) (\% m)$ is harder than inverting $y \equiv g^x (\% p)$.

Let $y \equiv h(x) (\% m)$. If inverting $y \equiv f(x) (\% p)$ is equivalent to inverting $y \equiv h(x) (\% p)$, where $p$ derives from $m$, inverting $y \equiv f(x) (\% m)$ is equivalent to $y \equiv h(x) (\% m)$. If inverting $y \equiv f(x) (\% p)$ is harder than inverting $y \equiv h(x) (\% p)$, then when inverting $y \equiv f(x) (\% p)$ is easier than or equivalent to inverting $y \equiv g^x (\% p)$, inverting $y \equiv f(x) (\% m)$ is equivalent to inverting $y \equiv h(x) (\% m)$, and when inverting $y \equiv f(x) (\% p)$ is harder than inverting $y \equiv g^x (\% p)$, inverting $y \equiv f(x) (\% m)$ is harder than inverting $y \equiv h(x) (\% m)$.

Fourthly, on Assumption 1, if an inversion problem is harder than DLP, it is also harder than IFP.

In light of the above four points, it is easy to understand that we only need to consider AGR over a discrete interval whose upper bound as a modulus is a prime.

### 2.4.4 Difference between Two Types of AGR

In asymptotic analysis, for AGR over a continuous interval, there is

$$\lim_{x \to \infty} (\log^k x / x) = 0, \text{ or } \lim_{x \to \infty} (x / \log^k x) = \infty,$$

where $k \geq 1$, and the unique logarithmic base $g$ is a rational, or a real with a finite precision.

On the other hand, for AGR over a discrete interval, it is possible that there is

$$\lim_{x \to \phi^k(p) \,\&\, \phi^k(p) \to \infty} (\log^k x / x) = 1, \text{ or } \lim_{x \to \phi^k(p) \,\&\, \phi^k(p) \to \infty} (x / \log^k x) = 1,$$

and moreover it is almost impossible that there is $g_0 = \ldots = g_k$, where $k \geq 1$.

In computation, for $y = x^n$, $y = g^x$, and $y = x^x$ with $x, n, g$ being positive rationals, each computing $y$

from known $x$ can be done in time relevant to a cubic polynomial or seeking an algebraic irrational.

On the other hand, for $y \equiv x^n$ (% $p$), $y \equiv g^x$ (% $p$), and $y \equiv x^x$ (% $p$) with $x$, $n$, $g$ being positive integers, each computing $y$ from known $x$ can be done in cubic polynomial time.

Again observe an example of inversion problems with $y$ known.

Let $y = f(x) = x^n$, and $y = h(x) = g^x$ with $x \in (0, \infty)$. Then $f'(x) = nx^{n-1}$, $h'(x) = g^x \ln g$, and further

$$\lim_{x \to \infty} f'(x) / h'(x) = \lim_{x \to \infty} \log(nx^{n-1}) / \log(g^x \ln g)$$
$$= \lim_{x \to \infty} (\log n + (n-1)\log x) / (x + \log \ln g)$$
$$= 0,$$

which indicates that seeking the transcendental irrational $\log y$ is generally harder than seeking the algebraic irrational $y^{1/n}$ in computational complexity on condition that approximations based on a finite precision are not considered.

Clearly, the larger the precision is, the closer $x$ is to $y^{1/n}$. However, when the precision is infinite, or the bit-length of the precision is greater than $2^{80}$, seeking the irrational $y^{1/n}$ is presently infeasible in polynomial time.

## 3  Results from AGR Are Compatible with Those from PTR

It is well understood that a cryptosystem is usually constructed over the prime field $\mathbb{GF}(p)$ or the multiplicative group $\mathbb{Z}_p^*$, and thus in section 3, 4, and 5, the application of AGR over a discrete interval will be discussed.

Note that for the convenience, in the following discussion, the tendency $x \to \phi^k(p)$ & $\phi^k(p) \to \infty$, $x^{-1} \to \phi^k(p)$ & $\phi^k(p) \to \infty$, or $\log x \to \phi^k(p)$ & $\phi^k(p) \to \infty$ is usually omitted in a limit expression.

Especially, it should be noted that in section 3, 4, and 5, because $\lim \log^k f(x) / \log^k h(x)$ is equivalent to $\lim \log^k f'(x) / \log^k h'(x)$ when middle infinitesimals are neglected, we substitute the latter with the former in the asymptotic analysis. That $\lim \log^k f(x) / \log^k h(x)$ is equivalent to $\lim \log^k f'(x) / \log^k h'(x)$ indicates that if $\lim \log^k f'(x) / \log^k h'(x)$ is a constant, infinitesimal, or infinity, $\lim \log^k f(x) / \log^k h(x)$ is also a constant, infinitesimal, or infinity.

### 3.1  Comparison between Inverting $y \equiv x^x$ (% $p$) and Inverting $y \equiv g^x$ (% $p$)

***Definition 7:*** Let $p$ be a prime. Seeking $x < p$ from $y \equiv x^x$ (% $p$) is called the transcendental (discrete) logarithm problem, in brief TLP.

#### 3.1.1  Proof of $\bar{I}(y \equiv g^x$ (% $p$)) $\leq_P \bar{I}(y \equiv x^x$ (% $p$)) by PTR

Essentially, we need to prove that
$$\bar{I}(y \equiv g^x \text{ (\% } p)) \leq_P \bar{I}(y \equiv (gx)^x \text{ (\% } p)) =_P \bar{I}(y \equiv x^x \text{ (\% } p)).$$

*Proof:*

Firstly, suppose that $g \in \mathbb{Z}_p^*$ is a generator coprime to $p - 1$. Notice that such a supposition does not lose generality since $g$ may be selected in practice.

Again suppose that $y \in \mathbb{Z}_p^*$ is known, and we need to seek $x$ such that
$$y \equiv (gx)^x \text{ (\% } p).$$

Raising either side of the above equation to the $g$-th power gives
$$y^g \equiv (gx)^{gx} \text{ (\% } p).$$

Let
$$z \equiv y^g \text{ (\% } p), \text{ and } w = gx,$$
where the latter is not a congruence, then
$$z \equiv w^w \text{ (\% } p).$$

Suppose that $\bar{O}_S(y, p, Q)$ is an oracle on solving $y \equiv x^x$ (% $p$) for $x$.

Its input parameters are $y$, $p$ and $Q$, where $Q$ is the set of all potential values of $x$, $p$ is a prime modulus, and $y \in [1, p-1]$.

Its output is $x \in Q$ (each of solutions), or 0 (no solution).

Let $Q_1 = \{1, 2, ..., p-1\}$, and $Q_2 = \{1g, 2g, ..., (p-1)g\}$.

Clearly, by calling $\bar{O}_S(y, p, Q_1)$, $y \equiv x^x$ (% $p$) is solved for $x$.

It is easily observed that between the limited sets $Q_1$ and $Q_2$, there is a linear bijection
$$\Gamma: Q_1 \rightarrow Q_2, \Gamma(a) = ga,$$
which means that the set $Q_1$ is equivalent to the set $Q_2$ [14]. Hence, substituting $Q_1$ with $Q_2$ will not increase the running time of $\bar{O}_S$.

Similarly, by calling $\bar{O}_S(z, p, Q_2)$, $z \equiv w^w$ (% $p$) is solved for $w$, namely all the satisfactory values of $w$ are obtained.

Further, $x \equiv wg^{-1}$ (% $p$), or $x \equiv wg^{-1}$ (% $p - 1$).

Therefore, in terms of definition 1, there is
$$\bar{I}(y \equiv (gx)^x \,(\% \, p)) \leq \bar{I}(y \equiv x^x \,(\% \, p)).$$

Namely the difficulty in inverting $y \equiv (gx)^x$ (% $p$) is not greater than that in inverting $y \equiv x^x$ (% $p$).

On the other hand, suppose that $\bar{O}_\S(y, g, p)$ is an oracle on solving $y \equiv (gx)^x$ (% $p$) for $x$.

Its input parameters are $y$, $g$, and $p$, where $p$ is a prime modulus, and $y, g \in [1, p - 1]$.

Its output is $x \in [1, p - 1]$ (each of solutions), or 0 (no solution).

Let $g = 1$.

By calling $\bar{O}_\S(y, 1, p)$, the solution $x$ to $y \equiv x^x$ (% $p$) will be obtained.

Therefore, in terms of definition 1, there is
$$\bar{I}(y \equiv x^x \,(\% \, p)) \leq_P \bar{I}(y \equiv (gx)^x \,(\% \, p)).$$

Further, in terms of definition 2, there is
$$\bar{I}(y \equiv x^x \,(\% \, p)) =_P \bar{I}(y \equiv (gx)^x \,(\% \, p)).$$

That is to say, the difficulty in inverting $y \equiv (gx)^x$ (% $p$) is equivalent to that in inverting $y \equiv x^x$ (% $p$).

Secondly, the congruence $y \equiv (gx)^x$ (% $p$) may be written as $y \equiv g^x x^x$ (% $p$), where $g$ is any generator.

Change $\bar{O}_\S(y, g, p)$ into $\bar{O}_\S(y, g, p, \hat{w})$, where $\hat{w} = 0$ or 1. Its structure is as follows:

S1: If $\hat{w} = 1$ and not $\exists x$ to $y \equiv g^x x^x$ (% $p$), return 'No'.

S2: If $\hat{w} = 1$,
   S2.1: find $y_1$, and compute $y_2$ by $y \equiv y_1 y_2$ (% $p$);
   S2.2: compute $x < p$ by $y_1 \equiv g^x$ (% $p$);
   S2.3: if $y_2 \neq x^x$ (% $p$), goto S2.1;
 else
   S2.4: compute $x < p$ by $y \equiv g^x$ (% $p$).

S3: Return $x$.

Clearly, by calling $\bar{O}_\S(y, g, p, 0)$, the solution $x$ to $y \equiv g^x$ (% $p$) will be obtained.

Therefore, still in terms of definition 1, there is
$$\bar{I}(y \equiv g^x \,(\% \, p)) \leq_P \bar{I}(y \equiv g^x x^x \,(\% \, p)).$$

In total, we have that
$$\bar{I}(y \equiv g^x \,(\% \, p)) \leq_P \bar{I}(y \equiv x^x \,(\% \, p)).$$

Namely inverting $y \equiv x^x$ (% $p$) is at least equivalent to inverting $y \equiv g^x$ (% $p$) in complexity. □

It is interesting whether inverting $y \equiv x^x$ (% $p$) is harder than inverting $y \equiv g^x$ (% $p$) or not.

Let $y \equiv g^t$ (% $p$), and $x \equiv g^u$ (% $p$), and then it seems that there is $g^t \equiv g^{ug^u}$ (% $p$).

However, due to $g^u$ (% $p$) $\neq g^u$ (% $p - 1$), $y \equiv x^x$ (% $p$) cannot be expressed as $t \equiv ug^u$ (% $p - 1$).

We can also understand that in the process of $x$ being sought from $y \equiv x^x$ (% $p$), it is inevitable that the middle value of $x$ will be beyond $p$ because modular multiplication, inverse, or power operations are inevitable.

Considering the middle value of $x$ beyond $p$, let
$$z_1 = x \,\%\, p, \text{ and } z_2 = x \,\%\, (p - 1),$$
where $z_1 < p$, and $z_2 < p - 1$.

Therewith, we have $x = z_1 + k_1 p = z_2 + k_2(p - 1)$ and $z_1 = (z_2 - k_2)$ % $p$, where $k_1, k_2 \geq 0$ are two integers. Further, we have $y \equiv (g(z_2 - k_2))^{z_2}$ (% $p$), which indicates that due to $x$ (% $p$) $\neq x$ (% $p - 1$) with $x > p$, and $k_2$ being variable, the relation between $x$ (% $p - 1$) and $x$ (% $p$) is stochastic when $x$ changes in the interval $(1, p^p)$.

Accordingly, it is reasonable that letting $v \equiv g(z_2 - k_2)$ (% $p$), namely we may have $y \equiv v^{z_2}$ (% $p$).

If $v$ is a constant, inverting $y \equiv v^{z_2}$ (% $p$) is equivalent to DLP. However, $v$ will not be a constant forever, which inclines us to believe that inverting $y \equiv x^x$ (% $p$) is harder than inverting $y \equiv g^x$ (% $p$).

### 3.1.2 Proof of $\bar{I}(y \equiv x^x \,(\% \, p))$ Being Harder than $\bar{I}(y \equiv g^x \,(\% \, p))$ by AGR

The result from AGR with assumption 1 reflects the layering of the two complexities more detailedly than that from PTR.

i) Proof through two steps

*Proof:*

Let
$$\log f(x) \equiv \log y \equiv \log (g\,x)^x \equiv x(1 + \log x) \,(\% \, \phi(p)),$$
and
$$\log h(x) \equiv \log y \equiv \log x^x \equiv x \log x \,(\% \, \phi(p)).$$
Then,
$$\lim \log f'(x) / \log h'(x) \cong \lim \log f(x) / \log h(x)$$
$$= \lim (x(1 + \log x)) / (x \log x)$$
$$= \lim (1 + 1/\log x).$$

Obviously, when $\log x \to \phi(p)$ and $\phi(p) \to \infty$, there are $1/\log x \to 0$, and
$$\lim \log (g x)^x / \log x^x = 1,$$
namely $\lim \log((g x)^x)' / \log(x^x)'$ is a constant.

Therefore, inverting $y \equiv x^x \,(\% \, p)$ is equivalent to inverting $y \equiv (g\,x)^x \,(\% \, p)$ in complexity.

In addition, let
$$\log f(x) \equiv \log y \equiv \log (g\,x)^x \equiv x(1 + \log x) \,(\% \, \phi(p)),$$
and
$$\log h(x) \equiv \log y \equiv \log g^x \equiv x \,(\% \, \phi(p)).$$
Then,
$$\lim \log f'(x) / \log h'(x) \cong \lim \log f(x) / \log h(x)$$
$$= \lim (x(1 + \log x)) / x$$
$$= \lim (1 + \log x).$$

Obviously, when $\log x \to \phi(p)$ and $\phi(p) \to \infty$, there are $\log x \to \infty$, and
$$\lim \log (g x)^x / \log g^x = \infty,$$
namely $\lim \log ((g x)^x)' / \log (g^x)'$ is infinite.

Therefore, inverting $y \equiv (g x)^x \,(\% \, p)$ is harder than inverting $y \equiv g^x \,(\% \, p)$ in complexity.

In total, we have that inverting $y \equiv x^x \,(\% \, p)$ is harder than inverting $y \equiv g^x \,(\% \, p)$ in complexity. □

ii) Proof through one step

We can prove directly.

*Proof:*

Let
$$\log f(x) \equiv \log y \equiv \log x^x \equiv x \log x \,(\% \, \phi(p)),$$
and
$$\log h(x) \equiv \log y \equiv \log g^x \equiv x \,(\% \, \phi(p)).$$
Then,
$$\lim \log f'(x) / \log h'(x) \cong \lim \log f(x) / \log h(x)$$
$$= \lim (x \log x) / x$$
$$= \lim (\log x).$$

Obviously, when $\log x \to \phi(p)$ and $\phi(p) \to \infty$, there are $\log x \to \infty$, and
$$\lim \log x^x / \log g^x = \infty.$$
namely $\lim \log (x^x)' / \log (g^x)'$ is infinite.

Therefore, inverting $y \equiv x^x \,(\% \, p)$ is harder than inverting $y \equiv g^x \,(\% \, p)$ in complexity. □

## 3.2 Comparison between Inverting $y \equiv g^{x^n} \,(\% \, p)$ and Inverting $y \equiv g^x \,(\% \, p)$

We will illustrate further that the result from AGR is compatible with that from PTR. Still assume that $g = g_0$ is a generator of $\mathbb{Z}_p^*$.

### 3.2.1 Proof of $\bar{I}(y \equiv g^{x^n} \,(\% \, p)) =_P \bar{I}(y \equiv g^x \,(\% \, p))$ by PTR

It is well known that the (modular) root finding problem (RFP) may be converted into the discrete logarithm problem.
*Proof:*
Assume that $\bar{O}_d(y, p, g)$ is an oracle on solving $y \equiv g^x$ (% $p$) for $x$.

Its input parameters are $y$, $p$, and $g$, where $p$ is a prime modulus, and $g, y \in [1, p - 1]$. Its output is $x$.

Let $y \equiv g^{x^n}$ (% $p$) and $w \equiv x^n$ (% $p - 1$). Then, there is
$$y \equiv g^w \ (\% \ p).$$

By calling $\bar{O}_d(y, p, g)$, $w$ can be obtained.

Solving $w \equiv x^n$ (% $p - 1$) for $x$ may be converted to DLP. Further, by the Index-calculus algorithm for DLP [5][15], $x$ can be found out if it exists, which indicates that $\bar{I}(y \equiv g^w \ (\% \ p)) \leq_P \bar{I}(y \equiv g^x \ (\% \ p))$.

On the other hand, the expression $y \equiv g^w$ (% $p$) is substantially the same as $y \equiv g^x$ (% $p$).

Thus, the difficulty in inverting $y \equiv g^{x^n}$ (% $p$) is equivalent to that in inverting $y \equiv g^x$ (% $p$). □

### 3.2.2 Proof of $\bar{I}(y \equiv g^{x^n} \ (\% \ p))$ Being Equivalent to $\bar{I}(y \equiv g^x \ (\% \ p))$ by AGR

Reasonably, the logarithmic granularity of order 2 is selected. It should be stressed that $\log^2 f(x) \equiv \log_{g_1}(\log_g f(x)) \ (\% \ \phi^2(p))$, where $g_1$ is an element of $\mathbb{Z}^*_{\phi(\phi(p))}$, or a generator with $\mathbb{Z}^*_{\phi(\phi(p))}$ being cyclic.

*Proof:*
Let
$$\log^2 f(x) \equiv \log^2 y \equiv \log^2 g^{x^n} \equiv n \log x \ (\% \ \phi^2(p)),$$
and
$$\log^2 h(x) \equiv \log y \equiv \log^2 g^x \equiv \log x \ (\% \ \phi^2(p)).$$
Then,
$$\lim \log^2 f'(x) / \log^2 h'(x) \cong \lim \log^2 f(x) / \log^2 h(x)$$
$$= \lim (n \log x) / (\log x)$$
$$= n.$$

Obviously, when $\log x \to \phi^2(p)$ and $\phi^2(p) \to \infty$, there are $\log x \to \infty$, and
$$\lim \log^2 f(x) / \log^2 h(x) = n,$$
namely $\lim \log^2 (g^{x^n})' / \log^2 (g^x)'$ is a constant.

Thus, inverting $y \equiv g^{x^n}$ (% $p$) is equivalent to inverting $y \equiv g^x$ (% $p$) in complexity. □

## 4 Asymptotic Granularity Reduction of Other Several Inversion Problems

In what follows, we will compare $y \equiv x^n$ (% $p$) with $y \equiv g^x$ (% $p$), $y \equiv x^n$ (% $p$) with $y \equiv x^n + x + 1$ (% $p$), and $y \equiv g^{g_1^x}$ (% $p$) with $y \equiv g^x$ (% $p$) in complexity of inversion.

### 4.1 Comparison between Inverting $y \equiv x^n$ (% $p$) and Inverting $y \equiv g^x$ (% $p$)

In this section, the logarithmic granularity of order 1 is selected.
Let
$$\log f(x) \equiv \log y \equiv \log x^n \equiv n \log x \ (\% \ \phi(p)),$$
and
$$\log h(x) \equiv \log y \equiv \log g^x \equiv x \ (\% \ \phi(p)).$$
Then,
$$\lim \log f'(x) / \log h'(x) \cong \lim \log f(x) / \log h(x)$$
$$= \lim (n \log x) / x$$
$$= n \ (\lim \log x / x).$$

Owing to modular arithmetic, when $x \to \phi(p)$ and $\phi(p) \to \infty$, $\lim (\log x / x)$ vibrates between 0 and 1, namely $\lim (\log (x^n)' / \log (g^x)')$ does between 0 and a constant, which indicates that either inverting $y \equiv x^n$ (% $p$) is easier than inverting $y \equiv g^x$ (% $p$) — the case of gcd ($n, p - 1$) = 1 for example, or inverting $y \equiv x^n$ (% $p$) is computationally equivalent to inverting $y \equiv g^x$ (% $p$).

### 4.2 Comparison between Inverting $y \equiv x^n + x + 1$ (% $p$) and Inverting $y \equiv x^n$ (% $p$)

In this section, the logarithmic granularity of order 0 is selected.
Let

and
$$f(x) \equiv y \equiv x^n (x^{-n} + x^{-n+1} + 1) \ (\% \ p),$$

$$h(x) \equiv y \equiv x^n \ (\% \ p).$$

Then,
$$\lim f'(x) / h'(x) \cong \lim f(x) / h(x)$$
$$= \lim (x^n (x^{-n} + x^{-n+1} + 1)) / (x^n)$$
$$= \lim (x^{-n} + x^{-n+1} + 1).$$

When $x \to p$ and $p \to \infty$, if $x^{-n} \to 1$, there is $x^{-n+1} \to \infty$, namely $x^{-n} + x^{-n+1} + 1 \to \infty$, or if $x^{-n} \to \infty$, there is also $x^{-n} + x^{-n+1} + 1 \to \infty$. Thus when $x \to p$ and $p \to \infty$, there always exists $x^{-n} + x^{-n+1} + 1 \to \infty$, which indicates
$$\lim (x^n + x + 1)' / (x^n)' \to \infty.$$

Therefore, inverting $y \equiv x^n + x + 1 \ (\% \ p)$ is harder than inverting $y \equiv x^n \ (\% \ p)$ in complexity.

### 4.3 Comparison between Inverting $y \equiv g_1^{g_1^x} \ (\% \ p)$ and Inverting $y \equiv g^x \ (\% \ p)$

In this section, the logarithmic granularity of order 2 is selected.

Let
$$\log^2 f(x) \equiv \log^2 y$$
$$\equiv \log^2 g_1^{g_1^x} \equiv \log g_1^x \equiv x \ (\% \ \phi^2(p)),$$

and
$$\log^2 h(x) \equiv \log^2 y$$
$$\equiv \log^2 g^x \equiv \log x \ (\% \ \phi^2(p)).$$

Then,
$$\lim \log^2 f'(x) / \log^2 h'(x) \cong \lim \log^2 f(x) / \log^2 h(x)$$
$$= \lim (x / \log x).$$

Owing to modular arithmetic, when $x \to \phi^2(p)$ and $\phi^2(p) \to \infty$, $\lim (x / \log x)$ will vibrate between 1 and $\infty$, namely $\lim \log^2 f'(x) / \log^2 h'(x)$ will vibrate between a constant and infinity, which indicates that either $x$ to $y \equiv g_1^{g_1^x} \ (\% \ p)$ does not exist in $\mathbb{Z}^*_{\phi(\phi(p))}$, or inverting $y \equiv g_1^{g_1^x} \ (\% \ p)$ is computationally equivalent to inverting $y \equiv g^x \ (\% \ p)$.

## 5 AGR Is a Complement to PTR Meantime

In the ElGamal signature scheme [2], the discriminant of the verification algorithm is
$$y^a a^b \equiv g^M \ (\% \ p),$$
where $(a, b)$ is a signature, $(y, g)$ is a public key, and $M$ is a message.

Now, assume that $b$ is known, and a counterfeiter needs to seek $a$ from the discriminant, which is equivalent to solving the problem $y \equiv x^n g^x \ (\% \ p)$ for $x$.

Clearly, comparison between $y \equiv x^n g^x \ (\% \ p)$ and $y \equiv g^x \ (\% \ p)$ by PTR is arduous or infeasible so far because the expression $x^n g^x \ (\% \ p)$ cannot be adapted to the form of $g^x \ (\% \ p)$.

Alternatively, we try to employ AGR.

Let
$$\log f(x) \equiv \log y$$
$$\equiv \log x^n g^x \equiv x + n \log x \ (\% \ \phi(p)),$$

and
$$\log h(x) \equiv \log y$$
$$\equiv \log g^x \equiv x \ (\% \ \phi(p)).$$

Then,
$$\lim \log f'(x) / \log h'(x) \cong \lim \log f(x) / \log h(x)$$
$$= \lim (x + n \log x) / x$$
$$= \lim (1 + n (\log x) / x).$$

When $x \to \phi(p)$ and $\phi(p) \to \infty$, $\lim (\log x / x)$ varies between 0 and 1 owing to modular arithmetic, which indicates $\lim (1 + n \log x / x)$ varies between 1 and $n+1$, namely $\lim \log (x^n g^x)' / \log (g^x)'$ varies between two constants not less than 1, which is a constant interval.

Thus, difficulty in inverting $y \equiv x^n g^x \ (\% \ p)$ is generally equivalent to that in inverting $y \equiv g^x \ (\% \ p)$.

Similarly, inverting $y \equiv xg^x \;(\%\; p)$ is computationally equivalent to inverting $y \equiv g^x \;(\%\; p)$.

These two examples illustrate that AGR is a complement to PTR when PTR cannot be utilized expediently.

## 6   Conclusions

Provable security by polynomial time reduction is essentially asymptotic, and also provable security by asymptotic granularity reduction is essentially asymptotic. They are very helpful in increasing our confidence in the security of a cryptosystem, but when a security dominant parameter is comparatively small, there may possibly exist a few exceptions to theoretical results. In practice, we must consider the exact security of the cryptosystem with some specified parameters.

Results from PTR are not as strict as those from AGR. AGR illustrates that a significant difference in the complexities of two problems is existent, partitions the complexities of problems more detailedly, has higher operation efficiency, and is an extension and complement to PTR. However, some results from AGR do not indicate that P ≠ NP holds since assumption 1 on which AGR is based is not proven.

When a modulus is composite, IFP and the Chinese remainder theorem in number theory set up a bridge between AGR with a prime modulus and AGR with a composite modulus.

Perhaps there exist some computational problems ─ functions containing two or more variables for example, whose relative complexities cannot be proven through either PTR or AGR. In this case, we should resort to number theory, finite field theory, or other mathematic tools.

The problems which can be proven to be harder than RFP or DLP through AGR may be used for contriving new and secure asymmetric encryption or signature schemes.

## Acknowledgment

The authors would like to thank the Academicians Jiren Cai, Zhongyi Zhou, Jianhua Zheng, Changxiang Shen, Zhengyao Wei, Andrew C. Yao, Binxing Fang, Guangnan Ni, and Xicheng Lu for their important guidance, advice, and suggestions.

The authors also would like to thank the Professors Dingyi Pei, Jie Wang, Ronald L. Rivest, Moti Yung, Dingzhu Du, Mulan Liu, Huanguo Zhang, Dengguo Feng, Yixian Yang, Maozhi Xu, Hanliang Xu, Xuejia Lai, Yongfei Han, Qibin Zhai, Xiaoyun Wang, Yupu Hu, Rongquan Feng, Ping Luo, Jianfeng Ma, Dongdai Lin, Zhenfu Cao, Jiwu Jing, Haiwen Ou, Lusheng Chen, Wenbao Han, Chao Li, Lequan Min, Dake He, Hong Zhu, Bingru Yang, Zhiying Wang, Quanyuan Wu, and Zhichang Qi for their important counsel, suggestions, and corrections.